\documentclass[aps,reprint,twocolumn,footinbib,showpacs,superscriptaddress,longbibliography]{revtex4-1}
\usepackage{graphicx}% Include figure files
\usepackage{dcolumn}% Align table columns on decimal point
\usepackage{bm}% bold math
\usepackage{amssymb}
\usepackage{pifont}
\usepackage{amsmath}
\usepackage{times}

\begin{document}

\title{Low-Dimensional Dynamics of Populations of Pulse-Coupled Oscillators}
\author{Diego Paz\'o}
\affiliation{Instituto de F\'{\i}sica de Cantabria (IFCA), CSIC-Universidad de
Cantabria, 39005 Santander, Spain }
\author{Ernest Montbri\'o}
\affiliation{Department of Information and Communication Technologies,
Universitat Pompeu Fabra, 08018 Barcelona, Spain}

\date{\today}
\begin{abstract}
% In nature, 
% Vast 
Large communities of biological oscillators show a prevalent 
tendency to self-organize in time. This cooperative 
phenomenon inspired Winfree to formulate a mathematical model 
that originated the theory of macroscopic synchronization.
Despite its fundamental importance, a complete mathematical analysis 
of the model proposed by Winfree ---consisting 
of a large population of all-to-all pulse-coupled oscillators--- 
is still missing. 
Here we show that the dynamics of the Winfree model 
evolves into the so-called Ott-Antonsen manifold. 
This important property allows for an exact description
of this high-dimensional system in terms of a few macroscopic variables,
and the full investigation of its dynamics.
% in terms of biologically meaningful ingredients such as the shape of the 
%  pulse-like interactions, or the oscillators' phase response curves (PRCs).  
We find that 
% the duration of the pulses profoundly 
% influences the ensemble's propensity towards synchronization. 
% Specifically, 
brief pulses are capable of synchronizing 
heterogeneous ensembles which fail to synchronize with broad pulses,
specially for certain phase response curves. 
Finally, to further illustrate the potential of our results, we investigate 
the possibility of `chimera' states in populations of identical pulse-coupled oscillators. 
Chimeras are self-organized states in which 
the symmetry of a population is broken into a synchronous and an asynchronous part. 
Here we derive three ordinary differential equations describing 
two coupled populations, and 
uncover a variety of chimera states, including a 
new class with chaotic dynamics. 
% The possibility of such states in realistic oscillator networks is an issue that 
% is currently receiving great experimental and theoretical attention.

\end{abstract}
\maketitle

\section{Introduction}

In 1967, Arthur Winfree proposed the first mathematical model for the 
macroscopic synchronization observed in large populations of 
biological oscillators \cite{Win67}. These natural 
systems typically achieve synchrony via brief pulse-like signals emitted 
by the individual oscillators \cite{Win80,Str03}. Well-known 
examples of pulse-like interactions are the action potentials 
emitted by neurons and other cells
\cite{HH52}, the flashes of light emitted by fireflies \cite{BB68}, 
or the sound of hands in clapping audiences \cite{neda00}. 

Assuming weak coupling, Winfree 
exploited the separation of time scales to characterize the state of each 
oscillator solely by its phase variable $\theta$. Using 
analytical arguments and numerical simulations, Winfree discovered 
that a population of $N\gg 1$ all-to-all-coupled phase oscillators 
showed a phase transition to macroscopic synchronization at a 
critical value of the `homogeneity' of the population \cite{Win67,Win79}. 
Only a few years after Winfree's 
seminal paper, Kuramoto proposed a new phase model 
singularly amenable to mathematical analysis \cite{Kur75,Kur84}. 
The Kuramoto model captures in an elegant and simple way the transition 
to collective synchronization observed by Winfree, and  
rapidly became the canonical model to mathematically 
investigate synchronization phenomena \cite{Str00}. 

The Kuramoto model has motivated a great deal of theoretical work,  
and has been investigated under countless 
variations as well as used to model a number of physical, chemical, biological, 
social, and technological systems \cite{PRK01,Str03,MMZ04,ABP+05}. 
Yet, in 2008, Ott and Antonsen made a very important finding \cite{OA08}:  
Kuramoto-like models have solutions in a 
reduced invariant manifold. This result drastically simplifies
the task of investigating the collective dynamics of such 
systems.

However, despite their importance and generality, Kuramoto-like models 
---in which interactions are expressed by phase differences---   
are approximations of more realistic models 
such as the Winfree model, in the weak-coupling limit. 
Parameters of the original model do not usually have a simple mapping 
into the parameters of the Kuramoto-like model 
(see e.g.~\cite{WCS96,MP11p}).   
In contrast to Kuramoto-like models, the Winfree model incorporates explicit
pulse-like interactions and phase response curves (PRCs) \cite{Canavier,*prcbook} that are 
customarily obtained from experiments \cite{tateno07,*kralemann} or 
from biologically realistic conductance-based models \cite{Izh07}. 

So far, theoretical attempts to understand the dynamics of the Winfree model 
have had very limited success.
Beyond a valuable work
in 2001 \cite{AS01} and a few posterior studies \cite{quinn,*basnarkov},   
the lack of mathematical tractability of the model seems to be 
the drawback for its dissemination among scientists. 

In this paper we show that the Winfree model evolves into the so-called
Ott-Antonsen (OA) manifold \cite{OA08}. Under some circumstances ---which 
we make clear below---, this important property 
permits us to exactly describe this 
high-dimensional system by two ordinary differential 
equations. We exhaustively explore the effect of 
the PRC's shape and the pulse's width
on the collective dynamics of the Winfree model.
In general, the evolution of the Winfree model in the OA manifold, 
%The OA ansatz also 
opens the possibility of investigating phenomena that 
so far were analytically addressed
%only addressable 
using ``Kuramoto oscillators''. 
As an example, we uncover the existence of a variety 
of the so-called chimera states \cite{motter} 
in populations of ``Winfree oscillators''.

\section{The Winfree Model}

The Winfree model writes:
\begin{equation}
\dot\theta_i=\omega_i+Q(\theta_i) \frac{\varepsilon}{N}  \sum_{j=1}^N P(\theta_j),
\label{model} 
\end{equation}
where the overdot denotes derivative with respect to time,
the constant $\varepsilon$ controls the coupling strength, and the 
oscillators are labeled by $i=1,\ldots,N$.  
The presence of heterogeneity in the population is modeled via 
the natural frequencies $\omega_i$, which are drawn from a certain 
probability distribution $g(\omega)$ \cite{Win67,AS01,quinn,*basnarkov} 
(see also \cite{tsubo07}). 
The PRC function
$Q$, measures the degree 
of advance or delay of the phases when the oscillators are perturbed. 
We adopt here a PRC with a sinusoidal shape: 
\begin{equation}
Q(\theta)=\sigma-\sin(\theta+\beta).
\label{prc}
\end{equation}
A possible choice relating the offset $\sigma$ and 
the phase-lag parameter
$\beta$ is $\sigma=\sin\beta$, so that the PRC vanishes at $\theta=0$, as it is naturally assumed in 
neuronal modeling. If $\beta<\pi/2$ neuronal oscillators are referred to as Type-II, 
whereas $\beta=\pi/2$ corresponds to a Type-I neuronal oscillator 
\cite{Izh07,hansel95,ermentrout96,NHM+00,tateno07,tsubo07,goel02}. 

We complete the definition of system~\eqref{model} with the 
smooth pulse-like signal:
\begin{equation}
P(\theta)=a_n (1+\cos\theta)^n
\label{p}
\end{equation}
where the integer parameter $n\ge 1$ allows to control the width of the pulses. 
The normalizing
constant $a_n$ is chosen so that the integral of $P(\theta)$ equals $2\pi$.
Thus $a_1=1$, and for other values of $n$: $a_n=2^n (n!)^2 /(2n)!$.
Note also that the $n\to\infty$ limit of \eqref{p} is
$P(\theta)=2\pi\delta(\theta)$.

%%%%%%%%%%%%%%%%%%%%%%%%%%%%%%%%%%%%%%%%%%%%%%%%%%%%%%%%%%%%%%%%%%%%%%%

\section{Limit of weak coupling and nearly identical frequencies}

We begin our analysis of the  
model defined by Eqs.~\eqref{model}, \eqref{prc} and \eqref{p}.     
taking the limit of
small $\varepsilon$ and frequency diversity. 
Applying the classical perturbative averaging technique 
\cite{Kur84} we obtain:
\begin{equation}
\dot \theta_i^{({\rm av})}=\omega_i'+\left(\frac{n}{n+1}\right)  \frac{\varepsilon}{N}
\sum_{j=1}^N \sin\left[\theta_j^{({\rm av})}-\theta_i^{({\rm av})}-\beta\right]
\label{kuramoto}
\end{equation}
with $\omega_i'=\omega_i+ \varepsilon\sigma$.
Equation \eqref{kuramoto} is precisely the Kuramoto-Sakaguchi model \cite{SK86,*OW12}.
An interesting outcome of our derivation of Eq.~\eqref{kuramoto} is that 
the narrower the pulses (larger the $n$ values) in the original Winfree model,
the stronger is the effective coupling $\varepsilon_{\rm eff}=n\varepsilon/(n+1)$. 

In the case of a Lorentzian distribution of frequencies, 
\begin{equation}
g(\omega)= \frac{\Delta/\pi}{(\omega-\omega_0)^2+\Delta^2},
\label{lorentzian}
\end{equation}
a closed formula for the coupling at the emergence of 
a macroscopic cluster of synchronized oscillators exists \cite{SK86}:
\begin{equation}
\varepsilon_c^{({\rm av})}= \frac{2 \Delta}{\cos\beta} \left(\frac{n+1}{n}\right).
\label{kh_sk}
\end{equation}
Note that this linear dependence of $\varepsilon_c$ on $\Delta$ is an approximation.

\section{Low-dimensional dynamics of the Winfree model}

%%%%%%%%%%%%%%%%%%%%%%%%%%%%%%%%%
\begin{figure*}
\centerline{\includegraphics[width=170mm]{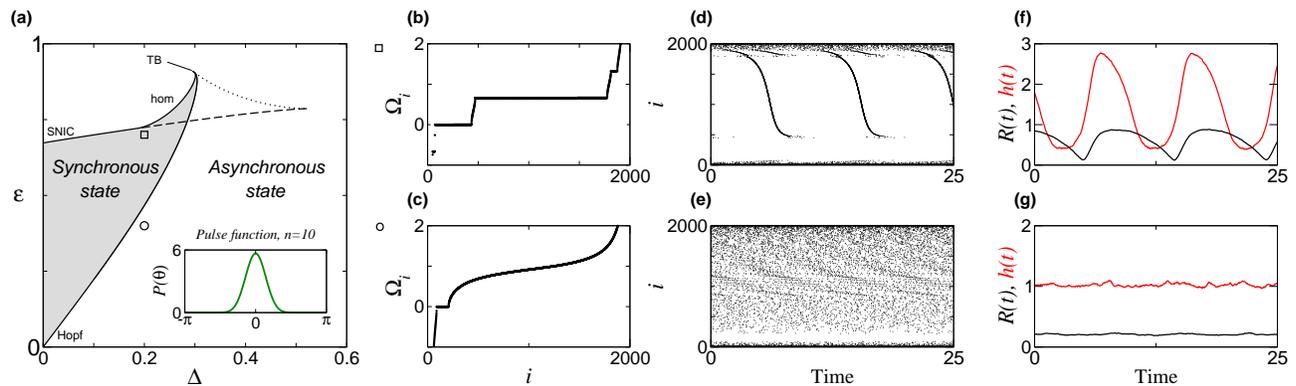}}
\caption{Color)  (a) Phase diagram of model \eqref{model}  obtained from the 
reduced Eqs.~\eqref{odes_polar},
with $\beta=\sigma=0$ and $n=10$.  
Inset: pulse-like function \eqref{p}.
Panels (b-g) show 
results obtained from the numerical integration of 
the Winfree model \eqref{model} 
with $N=2000$ oscillators and the natural frequencies
selected deterministically to represent the Lorentzian distribution:
$\omega_i=1+\Delta \tan(\pi/2 (2i-N-1)/(N+1))$, for $i=1,\dots,N$. 
Two different points ($\Box$, $\bigcirc$)  
corresponding to the synchronous (Top panels;~b,d,f), and asynchronous 
(Bottom panels;~c,e,g) states were chosen. (b,c): Coupling-modified 
frequencies: $\Omega_i=\lim_{t\to\infty}t^{-1}\int_0^t \dot\theta_i(t) dt$ 
versus oscillators' index $i$. 
Observe that in 
the synchronization region plateaus in $\Omega$ appear 
at a basic frequency and its integer multiples; other plateaus at rational
multiples of the basic frequency are absent due to the 
purely sinusoidal form of the PRC, see \cite{mirollo12}.
(d,e): raster plots (points depicted   
whenever $\theta_i=0$). (f,g): Time series of the modulus of
the Kuramoto 
order parameter $R(t)=|N^{-1} \sum_j e^{i\theta_j}|$ and the mean field $h(t)$.
}\label{fig1}
\end{figure*}
%%%%%%%%%%%%%%%%%%%%%%%%%%%%%%%

For the remainder of this paper, we analyze the Winfree model  
assuming neither weak coupling nor low frequency diversity. 
Our first key observation is that Eq.~\eqref{model}, 
with the PRC in Eq.~\eqref{prc}, belongs to a family of models
that can be written as:
\begin{equation}
\dot\theta_i(t)=\omega_i+ B(t) + \mathrm{Im}\left[H(t) e^{-i\theta_i(t)}\right].
\label{wbh}
\end{equation}
In our case, $B(t)=\varepsilon \sigma h(t)$ and
$H(t)= \varepsilon e^{-i\beta} h(t)$, with the mean field: 
\begin{equation}
h(t)=\frac{1}{N} \sum_{j=1}^N P(\theta_j(t)).
\label{mf}
\end{equation}
In the thermodynamic limit $N\to\infty$,
systems of type \eqref{wbh} have solutions in the reduced invariant 
manifold discovered by Ott and Antonsen \cite{OA08}, which
corresponds to a uniform distribution of certain
constants of motion at each value of $\omega$ \cite{PR11}.
For $B=0$, it has been proven \cite{OA09,OHA11} that,
provided the $\omega$'s are drawn from
a probability distribution function $g(\omega)$ which
is differentiable and well-behaved in a certain way (see \cite{OHA11} for details),
like Lorentzian or Gaussian functions, the dynamics of \eqref{wbh} converges to the
OA manifold. Remarkably, we have 
verified that the proof in \cite{OA09,OHA11} also holds for $B\ne0$. 
Hence, next we apply the OA ansatz with the certainty that it
captures the asymptotic dynamics of the model. 

Let $F(\theta|\omega,t)\, d\theta$ be the fraction
of oscillators with phases between $\theta$ and $\theta+d\theta$
and natural frequency $\omega$ at time $t$.
The dynamics of $F$ is governed 
by the continuity equation $\partial_t F= - \partial_\theta(\dot\theta F)$
since the number of oscillators is conserved. 
Using the OA ansatz
\begin{equation}
F(\theta|\omega,t)=\frac{1}{2\pi}
\left\{ 1+ \left[\sum_{m=1}^{\infty} \alpha(\omega,t)^m e^{im\theta}  + \mbox{c.c.}\right] \right\},
\label{oa}
\end{equation}
(where c.c.~stands for complex conjugate) 
we find that $\alpha(\omega,t)$ necessarily obeys:
\begin{equation}
\partial_t \alpha = -i (\omega+B) \alpha +  \frac{1}{2} ( H^*  -H \alpha^2).
\label{alpha}
\end{equation}
This is still an infinite set of equations
if the frequency distribution is continuous.
Fortunately, 
a drastic simplification is possible if $g(\omega)$ 
has a finite number of simple poles off the real axis
---like for the Lorentzian distribution \eqref{lorentzian}, see below.

For the analysis that follows, it is convenient to use the
generalized order parameters \cite{Dai96}:
\begin{equation}
Z_m(t)=\int_{-\infty}^\infty g(\omega) \int_0^{2\pi} F(\theta|\omega,t) e^{im\theta} d\theta d\omega
\label{zm},
\end{equation}
with $m\in\mathbb{N}$. Recalling the ansatz \eqref{oa}, and
noting that $\alpha$ admits an analytical continuation 
into the lower-half complex $\omega$-plane \cite{OA08},
we can evaluate \eqref{zm} applying the residue theorem.
Since the Lorentzian function \eqref{lorentzian}
has one simple pole $\omega^p=\omega_0-i\Delta$ inside the contour,
we obtain
that all order parameters depend on the value of $\alpha$ at the pole 
$Z_m(t)=[\alpha(\omega^p,t)^*]^m$.
The dynamics of the Kuramoto order parameter $Z_1\equiv R \, e^{i\Psi}$
is governed by two 
ordinary differential equations
(ODEs) obtained equating $\omega=\omega^p$ in \eqref{alpha}:
\begin{subequations}
 \label{odes_polar}
\begin{eqnarray}
\dot R &=& -\Delta R + \frac{\varepsilon h}{2}(1-R^2) \cos(\Psi+\beta) \label{ode1},\\
\dot \Psi&=& \omega_0 +\varepsilon  h\left[ \sigma  -\frac{1+R^2}{2R} \sin(\Psi+\beta) \right].
\label{ode2}
\end{eqnarray}
\end{subequations}
Remarkably, these two ODEs describe exactly
the Winfree model dynamics, 
irrespective of 
the particular interaction function $P(\theta)$.
In order to close Eq.~\eqref{odes_polar}, 
we consider $P(\theta)$ to be the pulse-like function in Eq.~\eqref{p}, and   
express the mean field \eqref{mf}
 in terms of $R$ and $\Psi$. For $n=1$ the result is trivial:
$h_1=1+R\cos\Psi$. For $n>1$, with the 
important observation
that the generalized order parameters are powers of the Kuramoto order parameter
$Z_m=Z_1^m$ \cite{PR11}, we obtain after some algebra:
\begin{equation}
h_n(R,\Psi)=1+ 2 (n!)^2 \sum_{k=1}^n \frac{R^k \cos(k\Psi)}{(n+k)! (n-k)!}.
\label{h}
\end{equation}

\begin{figure}
\centerline{\includegraphics[width=70mm]{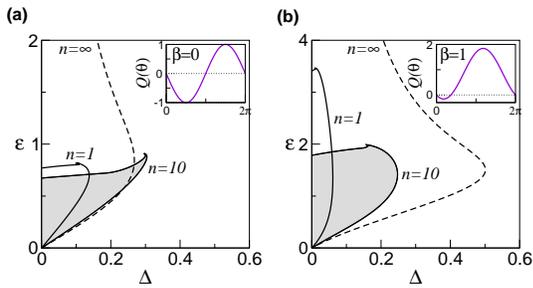}}
\caption{Color) Synchronization boundaries for PRCs with $\sigma=\sin\beta$,
and (a) $\beta=0$ and (b) $\beta=1$, and for pulse-like interactions 
\eqref{p} with $n=1$, 10 and $\infty$ (Dirac's delta). 
Insets: PRCs $Q(\theta)$, see Eq.~\eqref{prc}. 
The region of synchronization for $n=10$ appears shaded.
}
\label{betane0}
\end{figure}

Equation \eqref{odes_polar} cannot be solved analytically but the
loci of the bifurcations, where the qualitative behavior changes, can be easily
found by standard numerical continuation techniques.
We rescale $\Delta$, $\varepsilon$, and time by $\omega_0$, 
so that $\omega_0=1$ hereafter.
Additionally, we select $\sigma=\sin\beta$ 
---see the insets in Fig.~\ref{betane0}.
We have observed that the results are qualitatively the same independently 
of $n$ and $\beta$,
hence the phase diagram for $\beta=0$ and $n=10$ in Fig.~\ref{fig1}(a) 
accounts for all the phenomenology of the model.

The case $\beta=0$ was 
already studied in \cite{AS01} for a uniform distribution $g(\omega)$ 
obtaining a similar result, albeit some differences show up
due to the different support of the distributions. 
In the phase diagram of Fig.~\ref{fig1}(a), a Hopf bifurcation line emanates from
the origin ---with the slope predicted by Eq.~\eqref{kh_sk}--- limiting the 
shaded 
region of synchronization together with the other solid lines. 
In the synchronous state
a macroscopic cluster of oscillators rotates with the same coupling-modified
frequency $\Omega$, and as a result, the order parameter and the mean field oscillate;
see Figs.~\ref{fig1}(b,f).
Notice that, in addition, a cluster of oscillators with $\Omega=0$
and quivering near $\theta=0$ is present for all $\varepsilon>0$.
The region of synchronization is bounded at large values of $\varepsilon$
by a homoclinic (hom) and a saddle-node on the invariant cycle (SNIC) bifurcations.
The latter bifurcation line intercepts the $\varepsilon$-axis at 
$(n+1)^{n+1}/[a_n (2n+1)^{n+1/2}]$, {\em i.e.} $\varepsilon=0.6735\ldots$ for $n=10$.
In the phase diagram of Fig.~\ref{fig1}(a) we see that the Hopf line ends at a
Takens-Bogdanov (TB) point \cite{Kuznetsov}, which with two other (codimension-two) points
organize the region where Hopf and SNIC bifurcations meet; and this
conveys bistability between the Synchronous and the Asynchronous
states inside a small region bounded by the dashed (saddle-node bifurcation),
Hopf, and homoclinic lines.

After the preliminary introduction to the model dynamics,
we focus on the effect that
pulses' shape and oscillators' PRC have on the 
phase diagram of Fig.~\ref{fig1}(a). 
The boundaries in Fig.~2(a) for $n=1$ and 10 evidence 
that the region of synchronization enlarges as $n$ grows, as
suggested by Eq.~\eqref{kh_sk}. 
It is interesting to note that the coordinate $\varepsilon$ 
of the TB point diverges with $n$, 
while the $\varepsilon$ values of the SNIC line
decrease. As a result the region of bistability widens
as $n$ grows since the 
SNIC bifurcation at the $\varepsilon$-axis approaches
the finite value $\sqrt{\frac{e}{2\pi}}=0.6577\ldots$ as $n\to\infty$,
while the $\varepsilon$ coordinate of the TB point progressively grows.
The study of large $n$ values is difficult due
to the highly convoluted form of Eq.~\eqref{h}.
It is therefore useful from a mathematical perspective to consider the 
idealization $P(\theta)=2\pi\delta(\theta)$.
Using the trigonometric representation of the Dirac's delta function we obtain the mean field:
\begin{equation}
 h_{\infty}(R,\Psi)=\frac{1-R^2}{1-2R \cos\Psi + R^2}
\end{equation}
In this derivation the $n\to\infty$ limit
is taken after the $N\to\infty$ limit,
and therefore any subsequent result using $h_\infty$ is expected to be a truly asymptotic one
as $n$ grows provided $N$ is kept sufficiently large. On the contrary,
implementing instantaneous interactions ($n=\infty$)  with a finite population ($N<\infty$)
cannot fit in the theory since the mentioned limits do not commute
(this noncommutativity was studied in \cite{ZLP+07} for 
a model of leaky integrate-and-fire neurons).

Inserting $h_\infty$ in Eq.~\eqref{odes_polar} we obtain the boundaries\footnote{For $\beta=0$,
the boundary is $\varepsilon_c=\frac{1+5\Delta^2 \pm \sqrt{1-14\Delta^2+\Delta^4}}{6\Delta}$
with $\Delta\le 2-\sqrt{3}$.}
shown with dashed lines in Figs.~\ref{betane0}(a) and \ref{betane0}(b).
We see that for $\beta=0$ there is already a noticeable similarity between the regions of 
synchronization for $n=10$ and $n=\infty$. The main discrepancy is observed at high $\varepsilon$
values, which is not particularly interesting since, in any case, 
almost the whole population does not rotate in that region 
(see \cite{EK90} for a description of this effect).
As said above, as $n$ grows the TB point moves upwards, so that
the synchronization regions eventually match at the $n\to\infty$
limit (note nevertheless that the limit is somewhat singular because the bistability region disappears). 

For $\beta=1$, see Fig.~\ref{betane0}(b),
the difference between the results for $n=10$ and $n=\infty$
becomes apparent, and more tangible than what could be naively expected from
Eq.~\eqref{kh_sk}. 
In fact, the closer $\beta$ approaches to $\pi/2$, the more favorable is
a sharp $P(\theta)$ to achieve synchronization.
We claim this is a general statement, since
we have also observed it numerically with Gaussian $g(\omega)$.
It may be conjectured that the effectiveness of sharp spikes to 
achieve synchronization is one reason for their ubiquity in nature.

%%%%%%%%%%%%%%%%%%%%%%%%%%%%%%%%%%%%%%%%%%%%%%
%%%%%%%%%%%%%%%%%%%%%%%%%%%%%%%%%%%%%%%%%%%%

\section{Two coupled populations: Chimera States}

Finally, we aim to illustrate how our results permit 
to investigate problems that, so far, were only  
analytically addressed using the Kuramoto model. Recently, an interesting 
dynamical state, called chimera,
has been discovered in which identical oscillators
with identical connectivity self-organize into clusters  
with different synchronous behavior \cite{motter}. 
In this state complete synchronization of all the oscillators
is a stable solution and, therefore, the chimera state 
does not appear via a usual symmetry breaking mechanism\footnote{Note
that this is not the case of
the chimera-like state found in ensembles of leaky integrate-and-fire 
oscillators in S.~Olmi, A.~Politi, and A.~Torcini, 
Europhys. Lett. {\bf 92}, 60007 (2010).}. 
The simplest set-up capable of sustaining chimeras,
in both experimental \cite{tinsley12,martens} 
and numerical \cite{AMS+08,MKB04} realizations, consists of two 
coupled subpopulations $b=(1,2)$ of identical oscillators.  
Here, we consider a pulse-like coupling with
the positive constants $\mu$ and $\nu$ controlling intra- and 
inter-population interactions, respectively:
\begin{equation}
\dot\theta_i^{(b)}=1+Q(\theta_i^{(b)})  \left[\frac{\mu}{N_b} \sum_{j=1}^{N_b} P(\theta_j^{(b)})+
\frac{\nu}{N_{b'}} \sum_{j=1}^{N_{b'}} P(\theta_j^{(b')}) \right], \nonumber
 \end{equation}
with $b'=(2,1)$. Note that the equation for each  
subpopulation has the structure of Eq.~\eqref{wbh}
with $\omega_i=1$, $B\sigma^{-1}=H e^{i\beta}=\mu h^{(b)}+\nu h^{(b')}$,
and $h^{(b)}=\frac{1}{N_b} \sum_{j=1}^{N_b} P(\theta_j^{(b)})$.
In consequence, there is a solution in which each subsystem evolves into its own
OA manifold \eqref{oa}. The absence of diversity in the populations
makes the OA manifold to be neutrally stable \cite{PR08,PR11,WS94}.
Nevertheless, the OA manifold
becomes attracting as soon as a tiny amount of diversity is present \cite{Lai09}.
Thus, in some sense, the OA manifold is the ``skeleton'' of the
phase space, and it is legitimate to analyze the system with the OA ansatz.

The ODEs governing the dynamics of the order parameter of the $b$-th 
subpopulation $Z_1^{(b)}\equiv R_be^{i\Psi_b}$ ---cf. Eq.~\eqref{odes_polar}--- are:
\begin{eqnarray}
\dot R_b &=& \frac{\mu h^{(b)}+\nu h^{(b')}}{2} (1-R_b^2) \cos(\Psi_b+\beta),\\
\dot \Psi_b &=& 1 + \left[\mu h^{(b)}+\nu h^{(b')}\right]\left[\sigma- \frac{1+R_b^2}{2 R_b} \sin(\Psi_b+\beta) \right]. \nonumber
\end{eqnarray}
As we are interested in states where 
one subpopulation is fully synchronized, say the first one
($R_1=1$), the equation for $R_1$ disappears.
We obtain then a system of only three 
ODEs (for $\Psi_1$, $R_2$, and $\Psi_2$)
that makes possible to carry out 
an exhaustive exploration of the chimera states.

%%%%%%%%%%%%%%%%%%%%%%%%%%
\begin{figure}
\centerline{\includegraphics *[width=70mm]{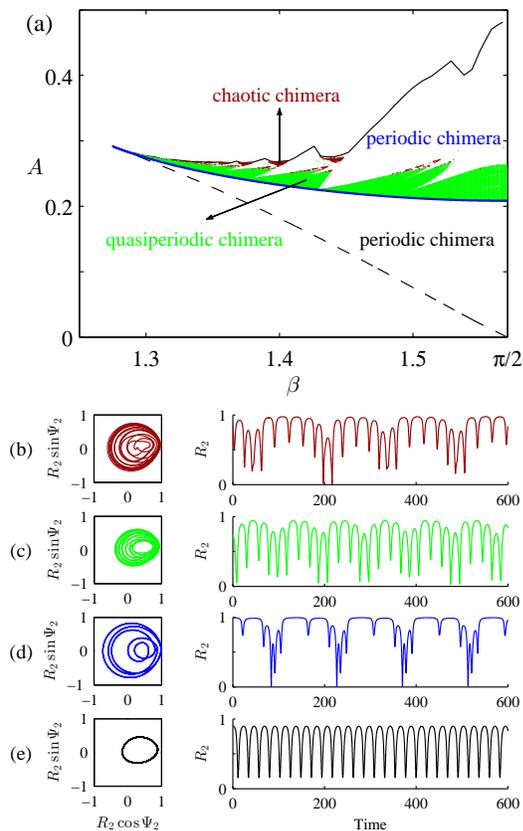}}
\caption{Color) (a) Location of different chimera types in the $(\beta,A)$ plane
for $\sigma=0$, $n=1$ and $S=0.5$
(similar results are obtained in a wide range of $\sigma$, $S$, and $n$).
Chimeras exist between the
dashed line (the locus of a saddle-node bifurcation of limit cycles),
and the solid line (corresponding to both boundary crisis \cite{AAIS,Ott} 
and saddle-node bifurcation of cycles). 
Above the thick (blue) line, quasiperiodic and chaotic chimeras are found
in the light (green) and dark (red) shaded regions, respectively.
(b-e) Trajectories projected onto $R_2 e^{i\Psi_2}$
and $R_2(t)$ for the parameter values
with distinct behaviors: $(\beta,A)=(1.4,0.265)$, $(1.42,0.24)$,
$(1.5,0.35)$, and $(1.45,0.19)$ from (b) to (e),
corresponding to chaotic, quasiperiodic, and periodic
chimera states above and below the thick (blue) line, respectively.}
\label{chimera}
\end{figure}
%%%%%%%%%%%%%%%%%%%%%%%%

It is convenient for the analysis to define two parameters: 
$A=(\mu-\nu)/(\mu+\nu)$ quantifying the imbalance
between intra- and inter-population interactions,
and $S=\mu+\nu$ quantifying the coupling strength.
Interestingly, in the limit of $\mu,\nu \to 0$, 
irrespective of the values of $\sigma$ and $n$,
the system reduces (via averaging)  to two 
ODEs for $R_2$ and $\psi=\Psi_1-\Psi_2$
identical to those in Eq.~(12) of Abrams et al.~\cite{AMS+08}
for oscillators of Kuramoto-Sakaguchi type.
Hence, for $S\to0$ we can borrow the results in \cite{AMS+08},
in particular, the existence of chimeras only for $\beta$ values not far from
$\pi/2$.
However, if $S$ is not small the system behaves as genuinely three-dimensional.
The structure of the phase diagram in Fig.~\ref{chimera}(a) is reminiscent of the one in
Fig.~4 of Ref.~\cite{AMS+08}, but now a much richer scenario emerges
due to the additional degree of freedom.
Above the Neimark-Sacker (or secondary Hopf) bifurcation, signaled by a thick (blue) line,
we find quasiperiodic chimeras, and the expected resonance tongues corresponding to
limit cycles on the surface of the invariant torus. 
As we move away from the Neimark-Sacker bifurcation the torus breaks down \cite{AAIS} 
and the resonances merge giving rise to an intricate set of bifurcations (not shown, 
see~\cite{kirk93}).
Perhaps the most remarkable consequence of the torus break-down  
is the existence of chaotic chimera states in the dark (red)
shaded region of the phase diagram in Fig.~\ref{chimera}(a). 
Figs.~\ref{chimera}(b-e) show
trajectories projected onto the $R_2 e^{i\Psi_2}$ plane
and time series $R_2(t)$ for specific values of $\beta$ and $A$.

\section{Conclusions}

The Winfree model describes a population of heterogeneous
limit cycle oscillators, which interact via pulse-like signals. 
Our most important finding is that the Winfree model 
with sinusoidal PRC, see Eq.~\eqref{prc}, belongs to a family of systems 
with the form of Eq.~\eqref{wbh}, and that such systems 
have asymptotic dynamics in a reduced space, called Ott-Antonsen manifold. 
This important property allows to  
exactly describe the dynamics of the Winfree model
with only two ODEs, Eq.~\eqref{odes_polar},
in the case of Lorentzian frequency distribution. 
The phase diagrams in Figs.~\ref{fig1} and \ref{betane0} permit to understand the 
effect
% on the Winfree model
of four parameters:
$\Delta$, $\varepsilon$, $\beta$, and $n$ controlling 
the spread of the natural frequencies, the coupling strength, 
the PRC, and the pulses' width, respectively.  
Interestingly, we find that brief pulses (large $n$ values) 
are capable of synchronizing heterogeneous ensembles which 
fail to synchronize with broad pulses. This 
feature of brief pulses is increasingly  enhanced as the PRC 
becomes more off-centered (increasing $\beta$), 
i.e.~as it approaches Type-I PRCs ---see Fig.~\ref{betane0}(b).  
It is worth noticing this property is not captured 
applying averaging, see Eq.~\eqref{kh_sk}, 
since the approximation \eqref{kuramoto} only holds at low values of 
coupling and frequency heterogeneity. 
Finally, the potential of our findings is illustrated uncovering
a variety of chimera states in networks of pulse-coupled oscillators, 
which include a new class of chimeras with chaotic dynamics. 

Our work suggests a number of future lines of research. For example, 
it would be interesting to investigate the 
dynamics of the Winfree model with more realistic 
ingredients such as time-delayed interactions or
pulse-like functions with coupling kinetics.
In addition, our theory can readily incorporate 
external fields and multimodal frequency distributions.    
All in all, we believe our results will foster theoretical advances
on the  collective dynamics of oscillators' systems, 
upgrading the mathematical basis of 
macroscopic synchronization beyond Kuramoto-like models.

\acknowledgments
We thank Juan M.~L\'opez for a critical reading of the manuscript,
Arkady Pikovsky for interesting discussions, and John Rinzel
for pointing us to Ref.~\cite{Win79}.
DP~acknowledges support by Cantabria International Campus,
and by MINECO (Spain) under a Ram\'on y Cajal fellowship. 
We acknowledge support by the Spanish research projects
No.~FIS2009-12964-C05-05 and No.~SAF2010-16085.

\emph{Note added}.---Recently, it came to our attention that
in parallel to our work, other authors have used the
OA ansatz to study ensembles of pulse-coupled 
theta neurons \cite{LBS13,SLB14}.

%\bibliography{bibliografia}
%Merlin.mbs v4.21 2009-07-09.
%

\end{document}